\documentclass[english,twocolumn,prb,a4paper,showpacs]{revtex4}
\usepackage[T1]{fontenc}
\usepackage[latin1]{inputenc}
\usepackage{amsmath}
\usepackage{graphicx}

\makeatletter

\providecommand{\tabularnewline}{\\}

\makeatletter



\makeatother

\usepackage{babel}
\makeatother

\begin{document}

\title{An ab-initio calculation of the core-level x-ray photoemission \\
 spectra -Fe $3s$- and $1s$-core levels-}

\author{Manabu Takahashi,$^{1}$ Jun-ichi Igarashi$^{2}$ and Noriaki Hamada$^{3}$}

\affiliation{$^{1}$Faculty of Engineering, Gunma Univ, Kiryu, Gunma 376-8515,
Japan}

\affiliation{$^{2}$Faculty of Science, Ibaraki Univ., Mito, Ibaraki 310-8512,
Japan}

\affiliation{$^{3}$Faculty of Science and Technology, Science Univ. of Tokyo,
Chiba 278-8510, Japan}

\begin{abstract}
We develop a method of the ab-initio calculation for the core-level
x-ray photoemission spectroscopy (XPS). By calculating one-electron
states in the presence of core-hole potential, excited states are
constructed by distributing electrons on these one-electron states.
The overlap integrals between the excited states and the ground state
are evaluated by using the wavefunctions with and without the core-hole
potential, and finally the XPS spectra are obtained. Implementing
the procedure, we obtain the spin-resolved $3s$-core XPS spectra
in ferromagnetic iron without any adjustable parameters, in good agreement
with the experiment. The spectral shapes are quite different for different
spin channels. We explain the behavior in terms of the difference
in the one-electron states screening the core-hole potential. 
\end{abstract}


\maketitle

\section{Introduction}

Electronic structure calculations based on the density functional
theory with the local approximation have provided valuable information
on the ground state properties.\cite{Kohn1965} Various attempts
beyond the local density approximation (LDA) have also been worked
out. These approaches have made great success on predicting accurate
electronic structures without introducing any adjustable parameters.\cite{Jones1989}
However, the LDA and its extensions have not been suited to treat
highly-excited states, and are usually limited to calculate the ground-state
properties. For example, although the x-ray photoemission spectroscopy
(XPS) from inner-core states is a powerful method to investigate the
dynamical response to the local perturbation associated with the photo-excited
core-hole, \cite{Davis1986} the corresponding spectra have not been
calculated from the first principles.

The purpose of this paper is to develop a method of the ab-initio
calculation for the inner-core level XPS spectra. It is known that
the spectra show asymmetric shape as a function of binding energy
near the threshold in metals. \cite{Doniach1970} This phenomenon
is related to the singularity in the response function near the Fermi
edge.\cite{Mahan1967,Anderson1967,Nozieres1969} The spectra sometimes
show extra peaks called satellites in systems of strongly-correlated
electrons. One example is in systems of rare-earth-metal compounds,
where the main and satellite peaks are explained as {}``well-screened\char`\"{}
and {}``poorly-screened\char`\"{} states, respectively. \cite{Kotani1973,Kotani1974}
Another notable example is a satellite around 6 eV separate from the
threshold on the $2p$-XPS in a ferromagnetic metal Ni.\cite{Hufner1975}
It arises from a combined effect of screening the core-hole potential
and the interaction between electrons. Feldkamp and Davis (FD) analyzed
the $2p$-XPS spectra in Ni by developing a numerical method on the
Hubbard model.\cite{Feldkamp1980} Evaluating the overlap between
the excited states and the ground state, they obtained the spectra
in good agreement with the experiment.

The method we develop in this paper is as an extension of the FD theory
to an ab-initio level. First, we carry out the band calculation within
the LDA to obtain the one-electron states in the ground state. We
use the augmented plane wave (APW) method instead of the Korringa-Kohn-Rostoker
(KKR) method, since we need the wavefunctions later. Next, we consider
a system of supercells with one core-hole per cell. The core hole
is introduced by removing an electron from the core level, which wavefunction
is assumed to be localized. In this system, we carry out the band
calculation to obtain the one-electron states. To guarantee the charge
neutrality, we add one extra electron in each supercell. To be exact,
there should exist only one core-hole in crystal for the XPS event.
In this situation, as known in the impurity problem, the local charge
neutrality is naturally satisfied according to the Friedel sum rule.
\cite{Friedel1958} The supercell method works better as increasing
the cell size. Note that not only the effect of core-hole potential
but also that of electron-electron interactions are taken into account
in the one-electron states thus obtained through the exchange-correlation
potential. Discretizing the momentum space into finite number of points,
we distribute electrons into the one-electron states to construct
the excited states. The lowest-energy state with core hole is given
by piling electrons into all one-electron levels up to the Fermi level.
One electron-hole pair, two electron-hole pairs, and so on, are created
from this core-hole state. Finally, we calculate the XPS spectra by
evaluating the overlaps between these excited states and the ground
states with the help of the one-electron wavefunctions.

We substantiate this scenario by applying the procedure to the $3s$-core
XPS in the ferromagnetic Fe metal.\cite{Acker1988,Hillebrecht1990,Bagus1994,Xu1995,See1995Fe,Lademan1997,Kamakura2006}
The spectra are resolved with respect to the spin of photoelectron
or equivalently the spin of core hole. Hereafter we call the majority
and minority spin as the up and down spin, respectively. The core
hole with up (down) spin indicates that one core electron with up
(down) spin is missing. The spectra show a strong dependence on the
spin;\cite{Xu1995} for the core hole with up spin, the spectral
intensity is distributed in a wide region of binding energy with a
clear satellite peak; for the core hole with down spin, the spectra
have a slightly asymmetric peak with no satellite peak. These features
indicate that the screening effect on the core hole with up spin is
quite different from that with down spin. So far, such spectra have
been analyzed using a single band Hubbard model, \cite{Kakehashi1984}
but it is too simple to compare quantitatively the experimental data.
\cite{Xu1995} We demonstrate that the present scheme works well
by comparing the calculated spectra with the experimental data.

Since the core hole has a spin-degrees of freedom, the exchange effect
plays an important role, which is different from the impurity potential
problem in dilute alloys. \cite{Kanamori1965,Terakura1971} We could
explain the strong spin dependence of the spectra in relation to $3d$-bands
changed by the core hole. For the core hole with up spin, the $3d$
electrons with down spin are strongly attracted by the core-hole potential,
resulting in an overscreening, while the $3d$ electrons with up spin
are little influenced by the core-hole potential because of the exchange
effect, resulting in an antiscreening due to the repulsion between
$3d$ electrons. The local spin moment is reversed at the core-hole
site. On the other hand, for the core hole with down spin, the $3d$
electrons with up spin are strongly attracted by the core-hole potential,
but the unoccupied $3d$ states available to the screening are limited,
resulting in an incomplete screening. The shortage of the screening
is compensated by the $3d$ electrons with down spin. The spin-dependent
XPS spectra are interpreted in terms of these changes in the $3d$
one-electron states.

Note that the $3s$ satellite is conventionally interpreted as a multiplet
splitting between the $3s$ and $3d$ electrons on the basis of the
atomic model.\cite{Bagus1994} However, the electronic structure
given in this paper is quite different from what the atomic model
predicts. The number of $3d$ electrons is increased nearly one at
the core-hole site in accordance with Friedel sum rule, and the
one-electron states are strongly modified from the ground state, and
even quasi-bound states appear.

Furthermore, we calculate the $1s$-core XPS spectra to clarify the
situation by comparison with the $3s$-core spectra. Since the $1s$
wavefunction is so localized that the exchange effect on the $3d$
one-electron states is expected to be relatively small. This is actually
the case, and we obtain the XPS spectra, which are less dependent
on the spin.

The present paper is organized as follows. In Sec. 2, we formulate
the XPS spectra with the ab-initio method. In Sec. 3, we present the
calculated XPS spectra for Fe $3s$- and $1s$-core in comparison
with the experiment. The last section is devoted to the concluding
remarks.

\section{Formulation}

\subsection{XPS spectra}

We consider the situation that a core electron is excited to a high
energy state with energy $\epsilon$ by absorbing the x ray with energy
$\omega_{q}$. Then the probability of finding a photoelectron with
energy $\epsilon$ may be proportional to \begin{align}
 & I_{\sigma}^{{\rm XPS}}(\omega_{q}-\epsilon)=\nonumber \\
 & \qquad2\pi|w|^{2}\sum_{f}|\langle f|s_{\sigma}|g\rangle|^{2}\delta(\omega_{q}+E_{g}-\epsilon-E_{f}),\label{eq:xps}\end{align}
 where $w$ represents the transition matrix element from the core
state localized at a particular site to the state of photoelectron.
The $s_{\sigma}$ is the annihilation operator of a relevant core
electron, which is assumed to have only spin $\sigma$ as the internal
degrees of freedom. State $|g\rangle$ represents the ground state
with energy $E_{g}$, while $|f\rangle$ represents the final state
with energy $E_{f}$. For the definition of $|f\rangle$, the photoelectron
is excluded. In the actual calculation, we replace the $\delta$-function
by the Lorentzian function with the full width of half maximum (FWHM)
$2\Gamma_{s}$ to taking account of the life-time broadening of the
core level. It should be noted here that we can specify the spin of
core hole by resolving the spin of photoelectron. Since the core-hole
potential induces the screening in the final state, the electronic
structure would become different from the ground state. Therefore,
the spectra would be spin-dependent in the magnetic system, due to
the exchange interaction.

\subsection{Overlap integrals}

The KKR method could solve a single core-hole problem in the final
state. Actually it is successfully applied to alloy problems with
combined to the coherent-potential approximation (CPA).\cite{Akai1982,Akai1998}
However, it may not be useful in the present problem, because we need
the information of wavefunctions. Instead, we consider a periodic
array of supercells containing one core-hole per cell, and calculate
the one-electron states by means of the augmented plane wave (APW)
method. The larger the unit cell size is, the better results are expected
to come out. Inner-core states, for example $1s$, $2s$, $2p$, and
$3s$ in Fe, are treated as localized states within a muffin-tin sphere,
so that we could directly specify the core hole site. To ensure the
charge neutrality, we assume $n_{e}+1$ band electrons per unit cell
instead of $n_{e}$ band electrons in the ground state. One additional
electron per unit cell would not cause large errors in evaluating
one-electron states in the limit of large unit cell size. We write
the resulting single particle states with energy eigenvalue $\epsilon_{n}({\bf k})$
as \begin{equation}
\psi_{n{\bf k}}({\bf r})=\frac{1}{\sqrt{N_{c}}}\sum_{j}\phi_{n{\bf k}}({\bf r}-{\bf R}_{j})\exp(i{\bf kR}_{j}),\end{equation}
 with \begin{equation}
\phi_{n{\bf k}}({\bf r}-{\bf R}_{j})=u_{n{\bf k}}({\bf r}){\rm e}^{i{\bf k}({\bf r}-{\bf R}_{j})},\end{equation}
 where $u_{n{\bf k}}({\bf r}-{\bf R}_{j})$ is a periodic function
with changing unit cells, and $j$ runs over $N_{c}$ unit cells.
We use these one-electron states as substitutes for the states under
a single core-hole. We distribute $N_{e}$ ($\equiv n_{e}N_{c}$)
band electrons on these states to construct excited states. In addition,
we carry out the band calculation in the absence of the core hole
with assuming $n_{e}$ band electrons per unit cell. The wavefunction
and eigenenergy are denoted as $\psi_{n{\bf k}}^{(0)}({\bf r})$ and
$\epsilon_{n}^{(0)}({\bf k})$, respectively. All the lowest $N_{e}$
levels are occupied in the ground state. Then, the matrix elements
connecting the ground and final states are expressed as \begin{equation}
\langle f|s_{\sigma}|i\rangle=\left|\begin{array}{cccc}
S_{1,1} & S_{2,1} & ..... & S_{N_{e},1}\\
S_{1,2} & S_{2,2} & ..... & S_{N_{e},2}\\
.... & .... & .... & ....\\
S_{1,N_{e}} & S_{2,N_{e}} & ...... & S_{N_{e},N_{e}}\end{array}\right|,\label{eq.overlap}\end{equation}
 with \begin{equation}
S_{i,i'}=\int\phi_{i}^{*}({\bf r})\phi_{i'}^{(0)}({\bf r}){\rm d}^{3}r,\end{equation}
 where the integral is carried out within a unit cell. Subscripts
$i=(n,{\bf k})$ and $i'=(n',{\bf k}')$ are running over occupied
states in the final state and in the ground state, respectively. The
corresponding energies are \begin{eqnarray}
E_{f} & = & \epsilon_{1}+\epsilon_{2}+\cdots+\epsilon_{N_{e}},\label{eq.ef}\\
E_{g} & = & \epsilon_{1}^{(0)}+\epsilon_{2}^{(0)}+\cdots+\epsilon_{N_{e}}^{(0)}.\label{eq.eg}\end{eqnarray}
 Substituting Eqs.~(\ref{eq.overlap}), (\ref{eq.ef}), and (\ref{eq.eg})
into Eq.~(\ref{eq:xps}), we obtain the XPS spectra.

\begin{figure}[t]
 \includegraphics[clip,scale=0.7]{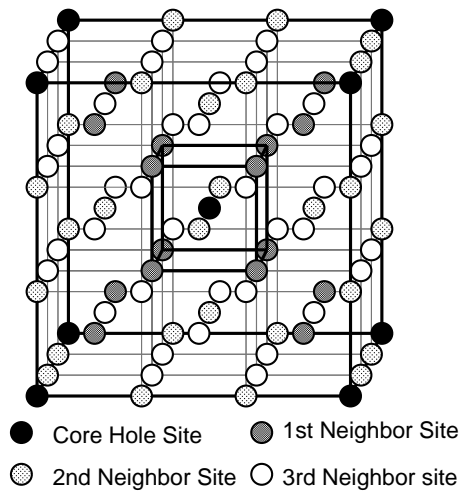}

\caption{Sketch of a supercell containing core holes in bcc Fe. \label{fig:unitcell}}

\end{figure}

\section{Calculated results}

\begin{figure}[h]
\begin{centering}
\includegraphics[scale=0.5]{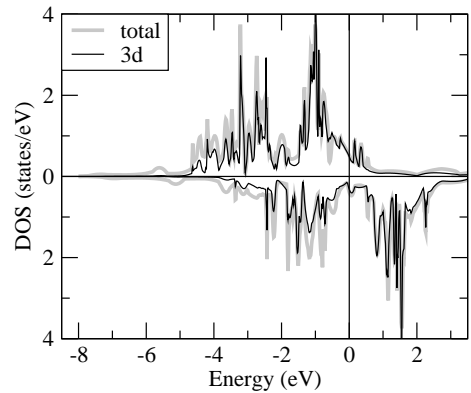} 
\par\end{centering}

\caption{DOS calculated in a system of supercells with no core hole in ferromagnetic
iron. The solid line represents the DOS projected onto the $d$-symmetry
within the muffin-tin sphere. \label{fig.DOSnohole}}

\end{figure}

We apply the above method to calculate the XPS spectra for the $3s$
and $1s$ cores in Fe metal. We use a supercell containing 27 iron
atoms, as illustrated in Fig.~\ref{fig:unitcell}. The first Brillouin
zone (BZ) for the supercells becomes much smaller than the original
first BZ. Accordingly, the energy bands in the absence of core hole
is folded. To demonstrate the reliability of our supercell calculation,
we show in Fig.~\ref{fig.DOSnohole} the DOS calculated in the supercell
system without core hole. The DOS resolved into the majority (up)
spin and minority (down) spin reproduces well the previous results.\cite{Morruzi1978}
In addition, the DOS is projected onto the $d$ symmetry ($d$-DOS)
within the muffin-tin sphere, which curve is close to the total DOS.
This indicates that the most part of the DOS arises from the $3d$
states.

\subsection{3s-core hole}

\subsubsection{one-electron states}

\begin{figure}[h]
\begin{centering}
\includegraphics[scale=0.9]{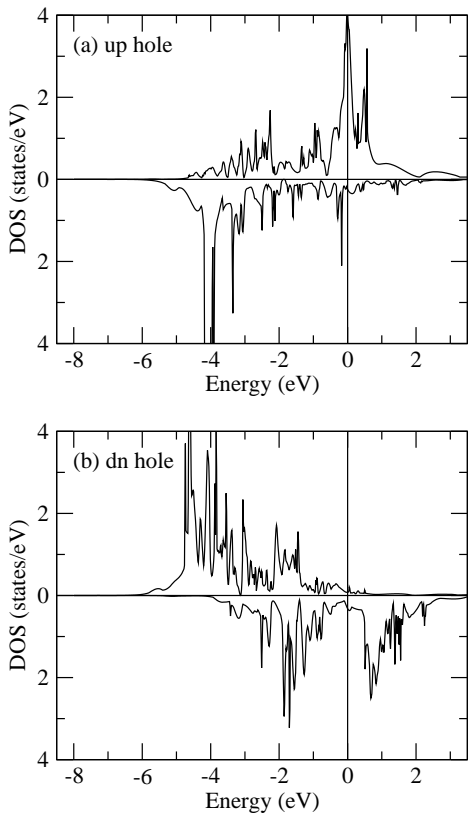} 
\par\end{centering}

\caption{The $d$-DOS at the site of $3s$-core hole with (a)up spin and (b)down
spin. \label{fig.DOS_3s}}

\end{figure}

How are the one-electron states modified by the core-hole potential?
We explain the change through the change in the $d$-DOS at the core-hole
site. Figure \ref{fig.DOS_3s} shows the $d$-DOS calculated in the
presence of the $3s$-core hole, and table \ref{table.1} lists the
screening charge $\Delta n_{d}$, which is defined by the change in
the occupied electron number caused by the core-hole potential within
the muffin-tin sphere at the core-hole site.

\begin{table}[tb]
\caption{Screening charge with respect to the $d$ symmetry within the muffin-tin
sphere on the $3s$-core-hole site. }

\label{table.1} \begin{tabular}{rrrr}
\hline 
 & $\Delta n_{d\uparrow}$  & $\Delta n_{d\downarrow}$  & $\Delta n_{d}$ (total)\tabularnewline
up spin hole  & -1.44  & 2.38  & 0.94 \tabularnewline
down spin hole  & 0.47  & 0.48  & 0.95 \tabularnewline
\hline
\end{tabular}
\end{table}

\subsubsection*{Up spin core-hole}

Figure \ref{fig.DOS_3s}(a) shows the $d$-DOS at the site of core
hole with up spin. By comparing with the $d$-DOS with no core hole,
we notice that the attractive core-hole potential is strongly working
on the $3d$ electrons with $down$ spin, pulling down the $d$-DOS.
A large weight develops around the bottom of the band. Even quasi-bound
states seem to be formed. It looks like an overscreening (the screening
charge reaches 2.38). On the other hand, the $3d$ electrons with
up spin are prevented from coming close to the core hole by the exchange
effect, thereby the attractive potential would be less effective.
The repulsion between $3d$ electrons pushes the $3d$ states upward,
resulting in a compensation of the overscreening (the screening charge
-1.44). As a total, the screening charge is 0.94. This indicates that
the screening is almost complete within the muffin-tin sphere on the
core-hole site.

\subsubsection*{Down spin core-hole}

Figure \ref{fig.DOS_3s}(b) shows the $d$-DOS at the site of the
core hole with down spin. The attractive potential is now strongly
working on the $up$ spin states, pulling down the $d$-DOS. Again
a large weight develops around the bottom of the band. The difference
is that the screening charge is limited to $0.47$ in the up spin
$3d$ states, even after nearly all occupied states are pulled down
below the Fermi level. Therefore, the screening has to be completed
by the down spin $3d$ states. Since the attractive potential is less
effective on the minority spin electrons due to the exchange effect,
there is no appreciable weight around the bottom of the band. The
screening is completed only through a moderate transfer of weight
from unoccupied levels to the occupied levels.

\subsubsection{XPS spectra}

\begin{figure}[h]
\begin{centering}
\includegraphics[scale=0.5]{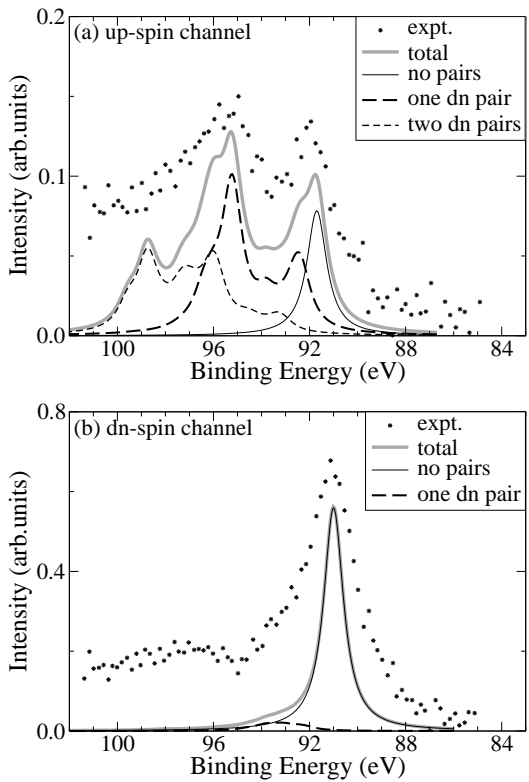} 
\par\end{centering}

\caption{XPS spectra as a function of binding energy for the $3s$-core hole
with (a) up spin and (b) down spin. The experimental data are taken
from Ref.~{[}\onlinecite{Xu1995}]. \label{fig.pes_3s}}

\end{figure}

We construct the excited states by distributing $N_{e}$ band electrons
on the one-electron states in the presence of the core hole. Since
the first BZ is quite small, we pick up only the $\Gamma$-point and
distribute $8\times27$ electrons on $3d$, $4s$, $4p$ bands. The
excited state with the lowest energy is given by piling one-electron
states up to the Fermi level. Then, picking up an occupied electron
from these states and placing it on a some unoccupied state, we create
one electron-hole pair. We create two electron-hole pairs in a similar
way. We calculate Eq.~(\ref{eq:xps}) with these excited states by
evaluating the Slater determinant in Eq.~(\ref{eq.overlap}) with
the use of the APW wavefunctions. Since the first BZ is quite small,
we take account of the states at the $\Gamma$-point only. Figure
\ref{fig.pes_3s} shows the calculated XPS spectra as a function of
the binding energy $\omega_{q}-\epsilon$ in comparison with the experiment.
\cite{Xu1995} The contributions from zero, one, and two electron-hole
pairs are separately shown. The calculated spectra reproduce well
the experimental data. In the following, we discuss the origin of
the spectra.

\subsubsection*{Up spin core-hole}

The excited state with the lowest energy (no electron-hole pair),
denoted as $|f_{0}\rangle$, has a finite overlap with the ground
state $|g\rangle$, giving rise to intensities on the threshold (broadened
by a Lorentzian function with FWHM $2\Gamma_{s}=2.0$ eV). The energy
of threshold is adjusted to coincide with the experimental value.
The overlap is given by $\langle f_{0}|s_{\uparrow}|g\rangle=A_{\uparrow}A_{\downarrow}$
with $A_{\uparrow}=0.980+i\,0.079$ and $A_{\downarrow}=-0.055-i\,0.352$,
where $A_{\uparrow}$ and $A_{\downarrow}$ represent the overlap
with respect to up spin electrons and with respect to down spin electrons,
respectively. In principle, such overlap tends to be zero with $N_{e}\to\infty$,
according to the Anderson orthogonality theorem.\cite{Anderson1967}
In such infinite systems, energy levels become continuous near the
Fermi level and thereby infinite numbers of electron-hole pairs could
be created with small excitation energies, leading to the so called
Fermi edge singularity in the XPS spectra. The finite contribution
obtained above arises from the discreteness of energy levels and could
be interpreted as the integral intensity of singular spectra near
the threshold, according to the model calculations to other systems.\cite{Kotani1974,Feldkamp1980}
Note that $|A_{\uparrow}|^{2}=0.966$, which is close to $1$, in
spite of the large modification of the local $3d$-DOS at the core-hole
site (see Fig.~\ref{fig.DOS_3s}(a)). This may be understood as follows;
since the $3d$ bands with up spin are almost occupied in the ground
state, the wavefunctions of the $3d$ bands in the presence of core
hole could be nearly represented by a unitary transform of the wavefunctions
of the occupied $3d$ electrons in the ground states. Therefore the
value of the Slater determinant in the up spin channel could not be
changed by the presence of the core hole. On the other hand, $|A_{\downarrow}|^{2}=0.127$,
suggesting that the wavefunctions of the occupied levels are considerably
modified from those of the ground state in the down spin channel,
which modifications are brought about by the mixing of the unoccupied
states of the ground state, because the $3d$ bands are only partially
occupied in the ground state. This change reduces the value of the
Slater determinant in the down spin channel.

For the excitation of one electron-hole pair, the contributions from
the up spin channel are negligibly small. Since the wavefunctions
of the unoccupied levels are almost orthogonal to the wavefunctions
of the $3d$ bands of the ground state, the Slater determinant corresponding
to the the occupation of such states would be almost vanishing. On
the other hand, the excitations of one electron-hole pair in the down
spin channel could give rise to a considerable contribution to the
spectra; each one-electron wavefunction of occupying electrons is
modified from those of the ground state by the mixing of the unoccupied
states, resulting in non-vanishing Slater determinants. As a result,
we obtain the intensities distributed with wide range of binding energy,
corresponding to various kind of one electron-hole pair. They consist
of two peaks; one is located around $\omega_{q}-\epsilon=92.5$ eV
and another is around $95$ eV. The latter peak corresponds to the
excitation from the quasi-bound states around the bottom of the $3d$
band in the down spin channel.

For the excitation of two electron-hole pairs, the contributions from
the up spin channel are extremely small by the same reason as for
the case of one electron-hole pair. The contributions arise from the
down spin channel; they consist of three peaks, one small peak around
$93.2$ eV, and large peaks around $96$ and $99$ eV. The excitation
of three electron-hole pairs gives rise to only minor contribution
to the spectra.

\subsubsection*{Down spin core-hole}

The excited state with the lowest energy (no electron-hole pair) has
an overlap with the ground state, $\langle f_{0}|s_{\downarrow}|g\rangle=A_{\uparrow}A_{\downarrow}$
with $A_{\uparrow}=0.708+i\,0.684$, $A_{\downarrow}=0.758+i\,0.576$.
Note that $|A_{\uparrow}|^{2}=0.969$, which is close to 1, although
the local $3d$-DOS is strongly modified from that in the ground state
(see Fig.~\ref{fig.DOS_3s}(b)). This means that, by the same reason
as explained in the up spin core-hole, the one-electron states in
the presence of core hole are nearly represented by a unitary transform
of occupied $3d$ states of the ground state, and thereby the corresponding
Slater determinant is close to 1. In the down-spin channel, $|A_{\downarrow}|^{2}=0.906$,
which is less than $|A_{\uparrow}|^{2}$ but close to 1. This simply
means that the modification of the one-electron states by the core-hole
potential is small, due to the exchange potential. Thereby we obtain
a large spectral weight around the threshold.

For the excitation of one electron-hole pair, we again have extremely
small contributions in the up spin channel because of the vanishing
overlap with the ground state. In the down spin channel, since the
modification of one-electron states by the core-hole potential is
small, the creation of electron-hole pair makes the corresponding
state nearly orthogonal to the ground state, and thereby small contributions
come out, as shown in Fig.~\ref{fig.pes_3s}.

\subsection{1s-core hole}

\subsubsection{one-electron states}

\begin{figure}[h]
\begin{centering}
\includegraphics[scale=0.9]{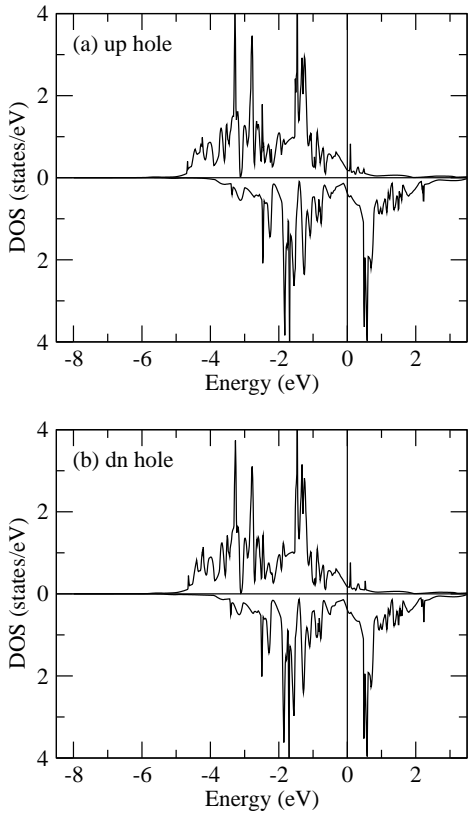} 
\par\end{centering}

\caption{The $d$-DOS at the site of $1s$-core hole with (a) up spin and (b)
down spin. \label{fig.DOS_1s}}

\end{figure}

Figure \ref{fig.DOS_1s} shows the $d$-DOS at the $1s$-core hole
site. It is a little modified from that with no core hole (Fig.~\ref{fig.DOSnohole}).
It depends little on the spin of the core hole, indicating that the
exchange effect is rather small. This is anticipated from the fact
that the $1s$ core is much localized around the origin than the $3s$
core. In addition, no quasi-bound state exists around the bottom of
the $3d$ bands. Table \ref{table.2} lists the screening charges
caused by the core-hole potential within the muffin-tin sphere at
the core-hole site. The screening charges are nearly independent of
the spin of core hole.

\begin{table}[tb]
\caption{Screening charge with the $d$ symmetry within the muffin-tin sphere
at the $1s$-core-hole site. }

\label{table.2} \begin{tabular}{rrrr}
\hline 
 & $\Delta n_{d\uparrow}$  & $\Delta n_{d\downarrow}$  & $\Delta n_{d}$ (total)\tabularnewline
up spin hole  & 0.35  & 0.74  & 1.09 \tabularnewline
down spin hole  & 0.36  & 0.73  & 1.09 \tabularnewline
\hline
\end{tabular}
\end{table}

\subsubsection{XPS spectra}

\begin{figure}
\begin{centering}
\includegraphics[scale=0.5]{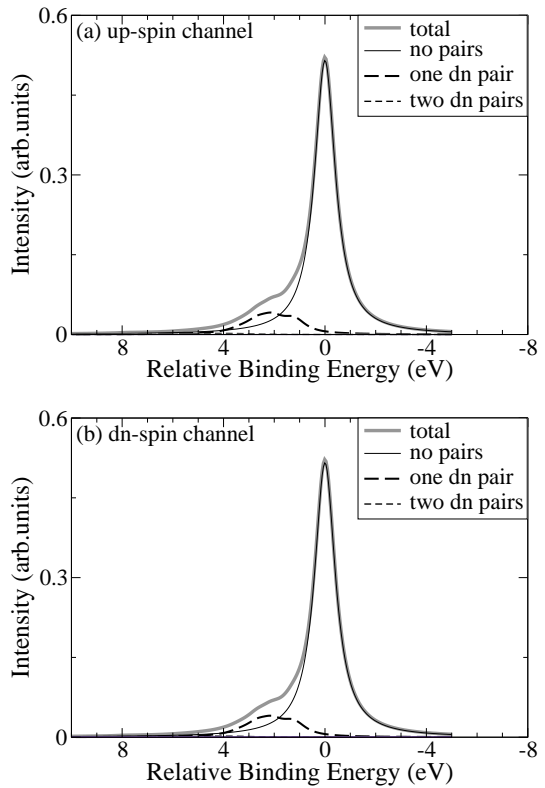} 
\par\end{centering}

\caption{XPS spectra as a function of binding energy for the $1s$-core hole
with (a) up spin and (b) down spin. \label{fig.pes_1s}}

\end{figure}

Figure \ref{fig.pes_1s} shows the XPS spectra for the core hole with
up spin and down spin, respectively. The spectra are broadened by
a Lorentzian function with FWHM $2\Gamma_{s}=2.0$ eV to take account
of the life-time broadening of the $1s$ core. The lowest energy state
(no electron-hole pair) in the presence of core hole has a finite
overlap with the ground state; $\langle f_{0}|s_{\uparrow}|g\rangle=A_{\uparrow}A_{\downarrow}$
with $A_{\uparrow}=0.836-i\,0.521$ ($A_{\uparrow}=-0.848+i\,0.501$),
$A_{\downarrow}=0.737+i\,0.538$ ($A_{\downarrow}=-0.708+i\,0.578$)
for the core hole with up (down) spin. Since $|A_{\uparrow}A_{\downarrow}|^{2}$
is $0.808$ ($0.810$) for the core hole with up (down) spin, most
intensities are concentrated near the threshold. Only small intensities
arise from the creation of electron-hole pair with no appreciate satellite.
Unfortunately, we don't know the experimental data to be compared
with.

\section{Concluding Remarks}

We have developed an ab-initio method to calculate the inner-core
XPS spectra. We have applied the method to the $3s$-core XPS in a
ferromagnetic metal Fe, where the spectra are resolved by the spin
of photoelectron. We have found that the spectral intensity distributes
in a wide range of binding energy with a clear satellite for the core
hole with up spin, while the intensity is concentrated near the threshold
with no satellite peak for the core hole with down spin, in good agreement
with the experiment. The origin of such behavior has been explained
in relation to the $3d$ band modified by the core-hole potential.

We have considered only the states on the $\Gamma$ point in the first
BZ to which electrons are distributed in the calculation of the XPS
spectra. Since the first BZ is reduced to a small size in a system
of supercells, this may not cause large errors. With increasing the
states to which electrons are distributed, one may expect that the
overlap between the lowest energy state in the presence of core hole
and the ground state would be reduced, and that the contributions
from electron-pair creation would increase near the threshold, leading
to an asymmetric peak near the threshold as a function of binding
energy. Such behavior has been demonstrated in numerical calculations
on finite size systems.\cite{Kotani1974,Feldkamp1980} On the other
hand, the structures with high binding-energy are expected to be only
a little influenced by such a refined treatment. In any case, to be
more quantitative, we need to increase $k$-points to which electrons
are distributed as well as to enlarge the size of supercells in the
calculation of the XPS spectra.

Acker et al. pointed out that there is poor correlation between the
satellite position and the magnetic moment on the Fe atom in various
alloys and compounds, and that the satellite splitting is observed
even in some Pauli paramagnets.\cite{Acker1988} We have interpreted
the satellite as arising from an excitation from quasi-bound states
to unoccupied states in one-electron states. At present it is not
clear how the position of quasi-bound state is correlated to the magnetic
moment in various systems and whether the satellite appears even in
Pauli paramagnets within the present approach. We need further studies
to clarify these points.

The present method could be extended to analyze XPS spectra from other
core holes, for example $2p$ core in Ni, and the x-ray absorption
spectra. \cite{Kruger2004} These extensions are left in future studies.

\begin{acknowledgments}
We would like to thank Professor T. Jo for valuable discussions. This
work was partly supported by Grant-in-Aid for Scientific Research
from the Ministry of Education, Culture, Sport, Science, and Technology,
Japan. 
\end{acknowledgments}
\bibliographystyle{apsrev} \bibliographystyle{apsrev}
\bibliography{Bibcore}

\end{document}